\documentclass[aps,prl,twocolumn,groupedaddress,showpacs]{revtex4-1}

\usepackage{braket} 
\usepackage{graphicx}
\usepackage{dcolumn}	% align numeric tables to decimal

\begin{document}

\title{Precision Measurement of Transition Matrix Elements via Light
Shift Cancellation}

\author{C.~D. Herold}
\email{cherold@umd.edu}
\author{V.~D. Vaidya}
\author{X. Li}
\author{S.~L. Rolston}
\author{J.~V. Porto}

\affiliation{Joint Quantum Institute, University of Maryland and NIST,
College Park, Maryland 20742}

\author{M.~S. Safronova}
\affiliation{Department of Physics and Astronomy, University of
Delaware, Newark, Delaware 19716}

\date{\today}

\begin{abstract}
  We present a method for accurate determination of atomic transition
  matrix elements at the $10^{-3}$ level. 
  Measurements of the ac Stark (light) shift
  around ``magic-zero'' wavelengths, where the light shift
  vanishes, provide precise constraints on the matrix elements.
  We make the first measurement of the 5$s$-6$p$ matrix 
  elements in rubidium by measuring the light shift 
  around the 421~nm and 423~nm zeros with a sequence of standing wave
  pulses.
  In conjunction with existing theoretical and experimental data,
  we find 0.3236(9)~$e a_0$ and 0.5230(8)~$e a_0$
  for the 5$s$-6$p_{1/2}$ and 5$s$-6$p_{3/2}$ elements, 
  respectively, an order of magnitude more accurate than the best
  theoretical values. This technique can provide needed, accurate matrix
  elements for many atoms, including those used in atomic 
  clocks, tests of fundamental symmetries, and quantum information.
\end{abstract}

% 03.75.Be : atom/neutron optics (cited in Saijun's paper)
% 37.10.Jk : atoms in optical lattices
% 37.10.Vz : mechanical effects of light on atoms
% 37.10.+k : atom interferometry techniques
% 67.85.-d : Ultracold gasses/ trapped gasses
%  67.85.Hj : BECs in optical potentials
% 32.70.Cs : Oscillator strengths, lifetimes, transition moments
% 32.10.Dk : electric and magnetic moments, polarizablilities
% 32.80.Qk : Coherent control of atomic interactions with photons

\maketitle

% ----------
%\section{Introduction} {{{1
Precise knowledge of atomic transition strengths is important in many
current areas of research, including the development of 
ultra-precise atomic clocks~\cite{Beloy2012,Sherman2012,PorLudBoy08},
studies of fundamental symmetries~\cite{PorBelDer09,VasSavSaf02} and
degenerate quantum gases~\cite{SafSafCla12},
quantum information~\cite{GorReyDal09,SafWal05},
plasma physics~\cite{SumBadOmu02}, and 
astrophysics~\cite{VerVerFer96,PicBlaTh11}. 
For example, one of the largest 
contributions to the uncertainty budget of atomic clocks is the
uncertainty in the blackbody radiation (BBR)
shift~\cite{Sherman2012,SafKozCla11}. 
The BBR shift is calculated from the difference in the
electric-dipole polarizabilities between the clock
states~\cite{PorDer06}, and its accuracy is currently limited by 
uncertainty in atomic transition matrix elements.
The dynamic correction to the BBR shift can also be determined
accurately if the relevant matrix elements are known ~\cite{Beloy2012}.
Studies of fundamental symmetries, such as atomic 
parity violation, need transition matrix elements to evaluate 
parity-violating amplitudes and vector transition
polarizabilities~\cite{PorBelDer09,VasSavSaf02}. 
Accurate AMO theory can also be indispensable to the design
and interpretation of experiments, where direct 
measurement of all relevant parameters is infeasible.
More complicated atoms, such as Er~\cite{AikFriMar12}, 
Dy~\cite{LuBurLev12}, and Ho~\cite{SafMol08} have recently become
of interest, and development of new theoretical methods must be
supported by the existence of high-precision experimental benchmarks.

Transition matrix elements can
be difficult to measure or calculate accurately. State-of-the art
theory predictions are often limited to a few percent uncertainty,
and the presently available experimental techniques typically
measure matrix elements accurately
only for one or two of the lowest transitions.
To date, the most accurate determinations of atomic transition matrix
elements are through excited state lifetime measurements or
photoassociation spectroscopy~\cite{Bouloufa2009}. 
The former is limited by uncertainty in branching ratios when
multiple decay paths exist. As a result, although the 6$p$
lifetime of rubidium was measured to 1\%, no matrix elements were
reported \cite{Gomez2004}.  The latter
requires species with purely long-range molecular excited states where
molecular theory is sufficiently well known to extract atomic
properties. 
In principle, transition matrix elements can be determined
from the ac polarizability $\alpha$ by measuring the 
ac Stark (light) shift of an atom exposed
to light of known intensity. However, unlike dc Stark shift measurements,
where the applied electric field can be determined
geometrically~\cite{Sherman2012}, 
it is difficult to accurately determine the optical
intensity, limiting the efficacy of this approach.
(Lifetime measurements avoid this calibration challenge by
using a well known field, the vacuum.) For atoms with spin-dependent
vector light shifts and two long-lived states, rf spectroscopy 
was used to accurately determine the \textit{ratio} of two vector light
shifts~\cite{Sherman2005,Sherman2008}, which when combined with
theory constrained matrix elements~\cite{Sahoo2009}. 
New techniques are needed to improve accuracy, extend the
range of measurable matrix elements, and provide benchmarks for
theory. 

In this Letter, we present a widely applicable method,
recently suggested in~\cite{Arora2011}, for
constraining matrix elements at the $10^{-3}$ level through direct
measurement of light shifts near ``magic-zero'' wavelengths
($\lambda_{\mathrm{zero}}$),  
where the combined light shift from all transitions
cancels~\cite{[{These are the ``tune-out'' wavelengths proposed for
species-selective traps by }]LeBlanc2007}. 
These $\lambda_{\mathrm{zero}}$ wavelengths are distinct from the
``magic'' wavelengths used in optical clocks, where the light shift is
identical for two states~\cite{Ye2008}.
We measure small light shifts through diffraction off a standing wave,
amplifying the diffracted populations by
constructively interfering the effect of up to $n_p=15$ pulses,
resulting in an $n_p^2$ enhancement in the diffracted fraction.
We make the first experimental 
determination of the rubidium $5s$-$6p_{1/2}$ and $5s$-$6p_{3/2}$ 
electric dipole matrix elements, to an accuracy of 0.3\%, 10 times
more accurate than the best theoretical values~\cite{Arora2011}. 
Implementation of our measurement technique could provide
much-needed data to constrain the BBR shift for optical lattice clocks.

Measuring light shifts near a $\lambda_{\mathrm{zero}}$ to constrain 
matrix elements has a number of advantages over the approaches
described earlier. It is insensitive to absolute calibration of the
applied optical field, requiring only a stable intensity.
Scalar light shifts are present for all atomic states, and
$\lambda_{\mathrm{zero}}$ points are found near every atomic
transition. That proximity increases sensitivity to the matrix element
of the nearby transition in addition to decreasing sensitivity 
to uncertainty in hard-to-calculate theoretical values.
Most importantly, ground (or metastable) state light shifts 
depend only on matrix elements directly to higher excited states, not on
the coupling of the excited states to other states. 
Light shift measurements are thus independent of the 
branching ratios that limit the lifetime approach.

\begin{figure}
  \includegraphics[scale=0.9]{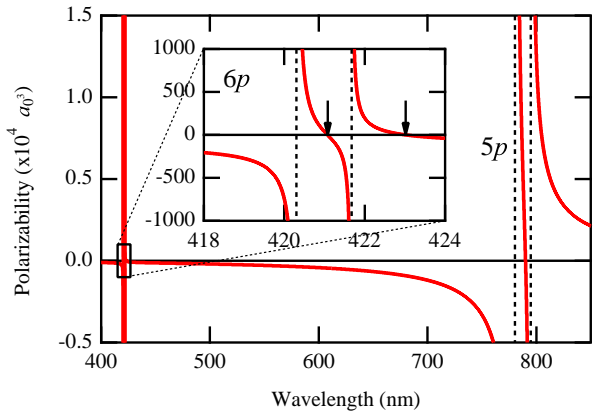}%0.9 2-col
  \caption{Calculated polarizability of Rb (atomic units of
	$a_0^3)$. The vertical
	dashed lines are at the $np$-5$s$ transitions. 
	Arrows indicate the positions of the 
	relevant $\lambda_{\mathrm{zero}}$ points. 
	\label{fig:RbPolar}}
\end{figure}

% ----------
%\section{Experiment} {{{1
%\subsection{Apparatus} {{{2
We used this technique to measure the transition 
matrix elements between the ground $5s_{1/2}$ and $6p_{1/2}$, $6p_{3/2}$ 
states in $^{87}$Rb by measuring the light shift as a function of $\lambda$
near the two $\lambda_{\mathrm{zero}}$ points neighboring the 6$p$
states (see the inset in Fig.~\ref{fig:RbPolar}).
We applied a sequence of standing wave (optical lattice) 
pulses to a Bose-Einstein condensate (BEC) containing 
$\simeq 3.5 \times 10^4$ $\ket{F=1, m_F=-1}$ 
atoms with no discernible thermal
fraction, produced in a hybrid optical and magnetic trap similar
to~\cite{Lin2009a}. At the end of the pulse sequence, we
suddenly turn off the trap and measure the atom population in each
diffracted order from an absorption image after 40~ms time of flight.
By intentionally using a small condensate, we can 
neglect atom-atom interactions, and limit 
the optical depth so that all orders in the image are unsaturated.
The measured population fractions, normalized to the total number in
each image, are insensitive to small fluctuations in total atom
number.

The lattice light is provided by a frequency-doubled diode laser, 
and is tunable between 419~nm and 424~nm.
We measure the wavelength to 50~fm (90~MHz) accuracy with
a wavemeter, calibrated to the known 5$s$-6$p$ transition
frequencies~\cite{Shiner2007}.
We form the lattice by reimaging a retroreflected beam on the atoms
with an incident power of 60-120~mW and a waist of $\simeq 110~\mu$m.
The lattice beam intensity varies only a few percent across the BEC, whose
Thomas-Fermi radius is 18~$\mu$m transverse to the beam.
Lattice  alignment is performed at 421.700~nm where the light
shift is reasonably large, $\simeq 7~E_R$ (where the recoil energy
$E_R=\hbar^2 k^2/(2M)$ for wavevector $k=2\pi/\lambda$, lattice laser
wavelength $\lambda$, and atomic mass $M$). The input beam is
carefully aligned to the BEC (with the retroreflected beam blocked) by
minimizing the transverse displacement of the BEC
induced by the beam. The retroreflected beam is then aligned by 
maximizing diffraction efficiency. Lattice depth measurements at the
5\% uncertainty level are consistent with no drift during a data set,
and the alignment is typically stable from day to day. 
%An acousto-optic modulator serves as a fast shutter to create
%the square diffraction pulses. 

To minimize wavelength-dependent steering of the laser,
we couple the light through an optical fiber
and focus the output of the fiber onto the BEC.  We measured the
residual steering after the fiber by monitoring the
position of the beam 1~m downstream from the
fiber tip. The observed deflection of the beam was $< 30~\mu$rad,
which corresponds to at worst a 1\% decrease in average beam intensity at the
BEC and is included in our uncertainty analysis.
Additionally, we monitor the power in
each pulse sequence with a photodiode to account for different
laser powers. 

Lattice polarization also affects the measured light shift. Linearly
polarized light has a purely scalar light shift, but elliptical 
polarization will have both vector and scalar light shifts for atoms
with $|m_F| > 0$~\cite{Deutsch2010a}, shifting the position of
$\lambda_{\mathrm{zero}}$. Although we used a Glan-laser polarizer
to establish a clean linear polarization,
analyzing the light after the chamber showed that the vacuum windows'
slight birefringence created $\simeq$ 1\% ellipticity.
To account for the possible systematic shift of
$\lambda_{\mathrm{zero}}$, we measured the light shift for two
orthogonal, linear input polarizations, defined as S and P with respect
to the plane of the last mirror before the chamber.
By symmetry, the induced ellipticity from
the window is opposite for the two input polarizations.
We calculated that averaging the S and P measurements cancels the
contribution of the vector light shift to a negligible 10~fm shift in
$\lambda_{\mathrm{zero}}$.

\begin{figure}
  \includegraphics[scale=0.7]{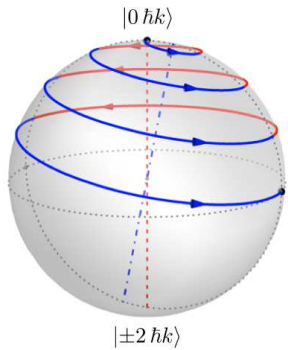}
	\caption{A Bloch sphere representation of the state vector's
	  evolution during a pulse sequence. The state vector starts in
	  $\ket{0\hbar k}$ (north pole). Lattice (free) evolution is shown
	  in blue (red). Alternating lattice and free evolutions with 
	  duration $h/(8E_R)$, the diffracted population is maximized 
	  as described in the text. This sequence provides 50\%
	  population in $\ket{\pm 2 \hbar k}$ after four pulses,
	  corresponding to a lattice depth of $\simeq 1 E_R$, much larger
	  than depths we actually measured.
	  \label{fig:BlochSphere}}
\end{figure}

%\subsection{Measurement Technique} {{{2
Optical lattice depths (light shifts) are commonly measured by
observing the population oscillations between 
momenta $2l\hbar k$, for integer $l$, as a function of lattice pulse
duration (outside the Raman-Nath regime).
For lattice depths below 1~$E_R$, the diffracted populations are
small, and the oscillation period saturates to $h/(4 E_R)$. 
At 0.05~$E_R$, less than $0.008$\%  of atoms are diffracted for 
a single lattice pulse.
In order to increase the signal, we employ a multi-pulse diffraction 
sequence that coherently adds the effect of each pulse, similar to 
quantum resonances in delta-kicked rotors~\cite{Izrailev1980,Moore1995}. 
For weak lattices where 
only orders with $|l| \leq 1$ are populated, a Bloch sphere picture
provides intuition~\cite{Wu2005} (Fig. \ref{fig:BlochSphere}). 
For initially stationary atoms, the $|l|=1$ orders are equivalent and can be
represented as a single state, which along with the $l=0$ state are the two
poles of the Bloch sphere. Free evolution corresponds to precession
about the vertical axis with period $h/(4E_R)$, called the Talbot 
time~\cite{Deng1999}. During a weak lattice
pulse, the precession axis is tilted by an angle proportional to the
lattice depth with essentially the same $h/(4E_R)$ period.
Alternating lattice and free evolution with duration $h/(8E_R)$
(one half precession period, $\simeq 9.75~\mu$s at 423~nm) efficiently 
increases the diffracted population (see Fig.~\ref{fig:BlochSphere}).

%In the weak lattice limit, the hamiltonian can be
%diagonalized analytically. While the lattice is on, the plane wave
%components are coherently mixed with time evolution governed by the
%band separation
%\begin{equation}
%  E_2-E_0 = 4 E_R \sqrt{1+\left( \frac{V_0}{4 E_R}\right)^2}\,,
%  \label{eq:bandsep}
%\end{equation}
%which is nearly identical to the free plane wave energy difference,
%$(2 \hbar k)^2/(2M) = 4 E_R$, for $V_0 \ll 1 E_R$. Thus, both
%evolutions have the same period, and the efficient pulse sequence is
%an even square wave with period $h/(4 E_R)$. 

\begin{figure}
  \includegraphics[scale=0.9]{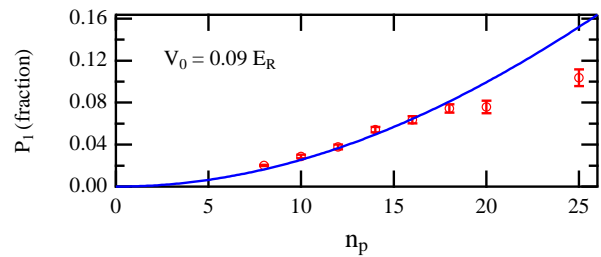}
  \caption{Diffracted population at fixed laser power and wavelength.
	As we vary the number of pulses, the diffracted population (our
	signal) initially increases as $n_p^2$. The solid line is 
	is the theoretical calculation with only $V_0$ as a fit parameter.
	For accurate light shift measurements, we only used $n_p \leq 15$.
    \label{fig:nPulses}}
\end{figure}

To accurately determine the lattice depths from the measured
populations, we numerically diagonalized the lattice Hamiltonian in
the plane wave basis including orders up to $|l|=3$. 
From this we calculate the first-order diffracted population 
fraction~$P_1$ as a function of pulse number $n_p$ and lattice depth
$V_0$ in units of $E_R$.
In the weakly diffracting limit ($n_p V_0 \ll 4$), 
$P_1 \propto (n_p V_0)^2 $, showing that multiple weak pulses are
equivalent to one pulse $n_p$ times stronger.
Inverting the exact numeric result for $P_1$, we extract the lattice depth 
$V_0(n_p,P_1)$ from the measured population fraction 
$P_1 = (N_{-1}+N_1)/N_{\mathrm{tot}}$,
with $N_l$ the number of atoms in the $l$th order. 
Fig.~\ref{fig:nPulses} shows $P_1$ versus $n_p$ at fixed $V_0$.
The agreement with theory is excellent for $n_p$ up to 18.
We attribute deviations at larger $n_p$ to the reduced overlap of the
diffracted components, which travel a significant fraction of the size
of the BEC, during the pulse sequence.

\begin{figure}
  \includegraphics{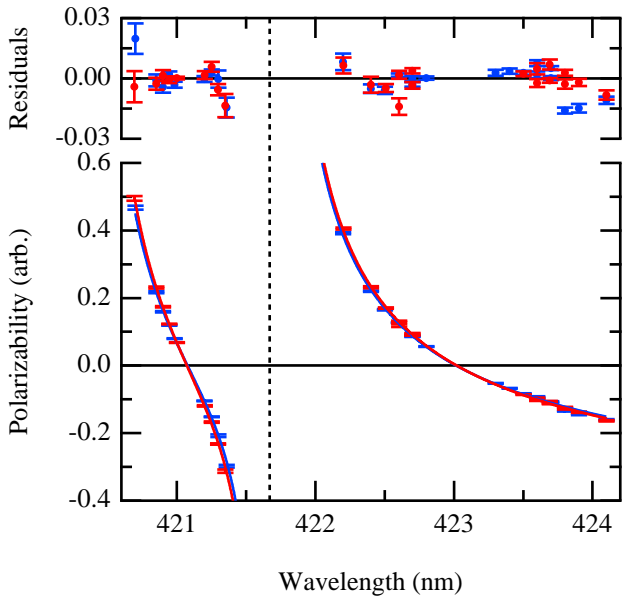}% scale=1 for 2-col (3.5'' wide)
  \caption{
	Measured polarizability (arb. units) versus wavelength 
	for two linear, orthogonal input polarizations S (red) and P (blue).
	Each point is an average of up to 20 measurements, 
	shown with 1$\sigma$ statistical error bars. 
	The solid lines (nearly indistinguishable) are fits 
	to the expected form (Eq.~\ref{eq:fit}) with the 6$p$ matrix elements
	as free parameters; reduced $\chi^2$ for the S and P fits is
	4.33 and 15.7, respectively. 
	Additionally, we allow a separate amplitude
	about each zero to account for different laser power. Fit
	residuals are shown at the top.
	\label{fig:data}}
\end{figure}

% ----------
%\section{Data} {{{1
To determine the polarizability, we extract the raw lattice depth  
$V_0(n_p,P_1)$ for each image, which is then divided by the measured
photodiode power for that pulse sequence.  Fig.~\ref{fig:data} shows 
the extracted polarizability as a function of $\lambda$ for four 
data sets: two orthogonal polarizations, 
S and P, near each $\lambda_{\mathrm{zero}}$. The lattice is aligned 
before each data set, and measurements 
after confirm  consistent alignment within a data set. 
The measurement is repeated up to 20 times at each 
wavelength~\footnote{Some points exhibited 
  decreased diffraction, correlated with the condensate position in each 
  image, due to residual condensate motion. 
  Thus we excluded the 22\% of images where the condensate was more than 
  15~$\mu$m from the average position for each data set. 
  The extracted matrix element is insensitive to the exclusion value.}, 
and the error bars in Fig.~\ref{fig:data} represent purely statistical 
uncertainty of 1$\sigma$. For both sets around the 421~nm zero, we used 
15 pulses, and for both sets around 423~nm we used 8 or 11 pulses.  

We fit the polarizability to the expression calculated in~\cite{Arora2011}:
\begin{equation}
  \alpha \propto \frac{1}{3} \sum_{n=5}^{8} \sum_{j=1/2}^{3/2}
  \frac{\left|d_{np_j}\right|^2\omega_{np_j}}{\omega^2-\omega_{np_j}^2}
  +C_{\mathrm{tail}}+C_{\mathrm{core}}\,,
  \label{eq:fit}
\end{equation}
where the $d_{np_j} \equiv \braket{np_j||D||5s}$ are reduced dipole
matrix elements and $\omega$ is the lattice laser frequency. 
The matrix elements $d_{6p_{1/2}}$ and $d_{6p_{3/2}}$ are fit
parameters, while $d_{5p_{1/2}}$ and $d_{5p_{3/2}}$ are fixed by the 
experimental lifetime measurement~\cite{Volz1996}. 
The 7$p$ and 8$p$ matrix elements and the 
contribution $C_{\mathrm{core}}$, which includes the polarizability of
the core electrons and their interaction with the valence electron,
are calculated in~\cite{Arora2011}. We used an improved calculation of 
$C_{\mathrm{tail}}$, the valence contribution for $n>8$, 
with reduced uncertainty (see supplemental material).

Fitting the data to Eq.~\ref{eq:fit} around a given $\lambda_\mathrm{zero}$
constrains the relationship among the parameters in the equation. For
example, the position of the $\lambda_\mathrm{zero}$ near 421~nm, which is
between the two $6p$ lines, depends strongly on the {\em ratio} of the
$6p$ matrix elements, $R_{6p}=d_{6p_{3/2}}/d_{6p_{1/2}}$, but more
weakly on their absolute value. On the other hand, the
position of the $\lambda_\mathrm{zero}$ near 423~nm, which is red-detuned of
both $6p$ lines, depends more on the average value of the $6p$
elements (in relation to the $5p$ elements). Fits around a
single $\lambda_\mathrm{zero}$ constrain $6p$ elements with respect to the
non-$6p$ parameters ($d_{np}$, $C_{\mathrm{tail}}$ and $C_{\mathrm{core}}$), 
and the extracted values are sensitive to their uncertainties.
Including data around
multiple $\lambda_\mathrm{zero}$ points further constrains the fits. For our
case, where the two $\lambda_\mathrm{zero}$'s are close to each other, the
effect of the non-$6p$ contributions is essentially identical for the
two zeros, so that simultaneous fits to both zeros accurately
determines $R_{6p}$, independent of the other parameters. With the
addition of sufficiently accurate theoretical and experimental values 
for the other $d_{np}$, $C_{\mathrm{tail}}$ and $C_{\mathrm{core}}$, we 
can also accurately determine absolute values 
for $d_{6p_{1/2}}$ and $d_{6p_{3/2}}$.

We simultaneously fit the 421~nm and 423~nm data sets for a 
given polarization. Due to possible differences
in alignment and total power between sets, we fit with an independent
amplitude, $A_{\lambda}$, around each $\lambda_{\mathrm{zero}}$
giving a total of four fit parameters: 
$A_{421}, A_{423}, d_{6p_{1/2}}, d_{6p_{3/2}}$.
The fit values are then averaged for the two polarizations.
(The small difference in extracted matrix elements for S and P
polarization is consistent with the 1\% ellipticity discussed
earlier.) Table \ref{tab:uncertainty} summarizes the 
contributions to our final uncertainty in our measured values 
for $R_{6p}$, $d_{6p_{1/2}}$ and $d_{6p_{3/2}}$.
The statistical uncertainty in $R_{6p}$, $d_{6p_{1/2}}$ and
$d_{6p_{3/2}}$ are roughly the same, of order 0.1\%. As expected, the
contribution of the uncertainties of the other fixed parameters in
Eq.~\ref{eq:fit} to $R_{6p}$ is negligible, providing a good
comparison with theory, which can predict ratios more accurately than
matrix elements (see supplemental material). 
The uncertainties in $d_{6p_{1/2}}$ and $d_{6p_{3/2}}$ do depend on
the other parameters, and the two largest contributions  are from the
theoretical uncertainty in the $np_{3/2}$ component of 
$C_{\mathrm{tail}}$ and the experimental uncertainty in
$d_{5p_{3/2}}$; these contribute at the 0.06\% level.
To account for potential lattice alignment drift, we simulated 
the effect of a 5\%
linear drift in the extracted polarizability across a data set, 
which resulted in an additional 0.2\% uncertainty in $d_{6p_{1/2}}$,
and 0.1\% in $R_{6p}$ and $d_{6p_{3/2}}$~\footnote{That reduced $\chi^2$
  for the fits in Fig.~\ref{fig:data} is larger than 1 is accounted for by
  the additional uncertainty in alignment drift; rescaling the
  statistical uncertainty to give reduced $\chi^2$ of 1 produces 
  similar error bars.}. 

\begin{table}
  \caption{Absolute uncertainty contributions (in $e a_0\times 10^{-4}$) for
  the $5s$-$6p$ matrix elements and their ratio ($\times 10^{-4}$). 
  Note the insensitivity of $R_{6p}$ to uncertainty in the fit parameters.
  Total uncertainty is summed in quadrature. Additionally, our $5s$-$6p$ 
  matrix elements are compared to the theoretical values (in $e a_0$).
  \label{tab:uncertainty}}
  \begin{ruledtabular}
    % D{...} from 'dcolumn' pkg
	\begin{tabular}{l D{.}{.}{-1} D{.}{.}{-1} D{.}{.}{-1}} 
	  Contribution 		% override D-alignment with \multicolumn
	  & \multicolumn{1}{c}{$\delta d_{6p_{1/2}}$}% ($\times 10^{-4}$)}	
	  & \multicolumn{1}{c}{$\delta d_{6p_{3/2}}$}% ($\times 10^{-4}$)}
	  & \multicolumn{1}{c}{$\delta R_{6p}$} \\% ($\times 10^-4$)} \\
	  \hline
	  statistical 		& 1.60 & 1.71 & 9.99 \\
      $d_{5p_{1/2}}$ 	& 0.84 & 1.35 & 0.005 \\
      $d_{7p_{1/2}}$ 	& 0.08 & 0.13 & 0.011 \\
      $d_{8p_{1/2}}$ 	& 0.02 & 0.04 & 0.002 \\
	  $np_{1/2}$ tail 	& 0.56 & 0.92 & 0.029 \\
	  $d_{5p_{3/2}}$ 	& 1.77 & 2.87 & 0.007 \\
	  $d_{7p_{3/2}}$ 	& 0.22	& 0.36 & 0.031 \\
	  $d_{8p_{3/2}}$ 	& 0.06	& 0.10 & 0.006 \\
	  $np_{3/2}$ tail 	& 2.01 & 3.28 & 0.104 \\
	  core 				& 1.25 & 2.05 & 0.065 \\
	  alignment drift	& 8.00 & 6.31 & 20.4 \\
	  Total	 		    & 8.74 & 8.29 & 22.7 \\
	  \hline
	  theoretical value~\cite{Arora2011} 
	  & \multicolumn{1}{D{.}{.}{6}}{0.325(9)} 
	  & \multicolumn{1}{D{.}{.}{6}}{0.528(13)} 
	  & \multicolumn{1}{D{.}{.}{6}}{1.624(7)} \\
	  our results		  
	  & \multicolumn{1}{D{.}{.}{6}}{0.3236(9)}
	  & \multicolumn{1}{D{.}{.}{6}}{0.5230(8)} 
      & \multicolumn{1}{D{.}{.}{6}}{1.616(2)} \\
    \end{tabular}
  \end{ruledtabular}
\end{table}

Our light shift cancellation measurement technique accurately determines 
the ratio of the 6$p$ matrix elements, with $R_{6p}=1.616(2)$.
The matrix elements are 
$d_{6p_{1/2}}=0.3236(9)$ and $d_{6p_{3/2}}=0.5230(8)$, in
excellent agreement with the theoretical values 
(Table~\ref{tab:uncertainty}). From these we determine values for the
$\lambda_{\mathrm{zero}}$ points: 421.075(3)~nm and 423.018(8)~nm.

% ----------
%\section{Conclusion} {{{1
We have presented the first experimental measurement of the 5s-6p
dipole transition matrix elements in Rb. 
Our technique of measuring the relative light shift near
$\lambda_{\mathrm{zero}}$ is applicable to many important
atoms. At a minimum, such measurements constrain matrix elements and
provide benchmarks needed for atomic theory. Absolute matrix element 
values can be bootstrapped from a single, well-known matrix element. 
Additionally, applying our measurement technique to 
metastable states, such as the $^3P_0$ clock state of Yb or Sr, 
could provide information about excited-excited state matrix elements.

\begin{acknowledgments}
We acknowledge useful discussions with G.K. Campbell, L.A. Orozco, and W.D.
Phillips. This work was partially supported by the NSF-PHY1104472 and the
ARO with funding from DARPA's OLE program and ARO Grant W911NF0910216
(DURIP).
The work of MSS was performed under the sponsorship of the US
Department of Commerce, NIST.
\end{acknowledgments}

% Create the reference section using BibTeX: {{{1 
\bibliography{../FinalBib}

% Insert supplemental info on new page
\newpage
\begin{widetext}

\begin{center}
  \textbf{{\large Supplemental Material}}
\end{center}

Calculation of the magic-zero (also called ``tune-out'') wavelengths
where the frequency-dependent
polarizability vanishes was discussed in detail in
Ref.~\cite{Arora2011}. Briefly,  the
frequency-dependent scalar polarizability, $\alpha_0(\omega)$, of an
alkali-metal atom in its
ground state $ v$ may be separated into a contribution from the core
electrons,
$\alpha_{\mathrm{core}}$, a core modification due to the valence
electron, $\alpha_{vc}$, and a
contribution from the valence electron, $\alpha_0^v(\omega)$. The
valence contribution to the
static ac polarizability is calculated using the sum-over-states
approach:
	\begin{equation}
	  \alpha_{0}^v(\omega)=\frac{2}{3(2j_v+1)}\sum_k\frac{{\left\langle
	   k\left\|D\right\|v\right\rangle}^2(E_k-E_v)}{
	  (E_k-E_v)^2-\omega^2}, \label{eq-1}
	\end{equation}
where $\left\langle k\left\|D\right\|v\right\rangle$ is the reduced
electric-dipole (E1) matrix
element. Because of the rapid convergence of the  sum  over
intermediate $k$ states in
Eq.~(\ref{eq-1}), we separate the valence state polarizability into
two parts,
  $\alpha_{\mathrm{main}}$,
containing the contributions from the few lowest $np$ states, and
the remainder,
$\alpha_{\mathrm{tail}}$. The experimental data \cite{Volz1996} are
used for $5s-5p$ matrix elements.
The $5s-6p$ matrix elements are determined in the present work.
For all other main  terms, we use the relativistic all-order
values~\cite{relsd} 
of the matrix elements and the experimental values of the energies
\cite{NIST1,NIST2}. In Ref.~\cite{Arora2011}, the contributions with
$n=5 - 8$ were calculated
accurately and tail was calculated in the Dirac-Hartree-Fock
approximations.  To decrease
uncertainty in the tail, we calculate $n= 9 - 12$ contributions using
all-order matrix elements and
experimental energies. The contributions to the frequency-dependent
polarizability  of the ground
state of Rb at $\lambda_{\mathrm{zero}}=421.07476$~nm and
$\lambda_{\mathrm{zero}}=423.01787$~nm magic-zero
wavelengths are given in Table~\ref{tab:supplement}. We list
$\alpha_{\mathrm{core}}$ and $\alpha_{vc}$
contributions together in the row ``Core''. The uncertainties in the
values of the magic-zero
wavelengths are taken to be the maximum difference between the central
value and the crossing of
the $\alpha_{{0}}\pm\delta\alpha_{0}$ with zero, where
$\delta\alpha_{0}$ is the uncertainty in the
ground state polarizability value at that wavelength. The resulting
values of
$\lambda_{\mathrm{zero}}$ 421.075(3)~nm and 423.018(8)~nm are in
excellent agreement with
421.08(3)~nm and 423.05(8)~nm predictions of Ref.~\cite{Arora2011},
respectively.
%%%%%%%%%%%%%%%%%%%
% This table is calculated values except for 5,6p
%%%%%%%%%%%%%%%%%%%
\begin{table*}[b]
  \caption{Contributions to the frequency-dependent polarizability 
	of the ground state of Rb at $\lambda_{\mathrm{zero}}=421.07476$~nm 
	and $\lambda_{\mathrm{zero}}=423.01787$~nm. Absolute values of electric
	dipole matrix elements are expressed in a.u. ($ea_0$), and the 
	corresponding energy differences are expressed in conventional 
	wavenumber units (cm$^{-1}$).
  \label{tab:supplement}}
  \begin{ruledtabular}
	\begin{tabular}{l D{.}{.}{1.7} D{.}{.}{5.2} D{.}{.}{4.8}D{.}{.}{4.8}}
	 Contribution      % override D-alignment with \multicolumn
	& \multicolumn{1}{c}{$|\braket{np_j||D||5s}|$}
	& \multicolumn{1}{c}{$E_{np_j}-E_{5s}$}
	& \multicolumn{1}{c}{$\alpha_{421.07476}$}
	& \multicolumn{1}{c}{$\alpha_{423.01787}$} \\
	\hline
	$5p_{1/2}$    & 4.231(3) & 12578.95 & -40.599(60) & -41.122(60) \\
	$6p_{1/2}$    & 0.3235(11) & 23715.08 & -113.608(771) & 50.840(346) \\
	$7p_{1/2}$    & 0.115(3) & 27835.02 & 0.128(6) & 0.125(6) \\
	$8p_{1/2}$    & 0.060(2) & 29835.00 & 0.024(2) & 0.023(2) \\
	\hline
	$9p_{1/2}$    & 0.037(3) & 30958.91 & 0.0080(11) & 0.0079(11) \\
	$10p_{1/2}$   & 0.026(2) & 31653.85 & 0.0036(5) & 0.0036(5) \\
	$11p_{1/2}$   & 0.020(1) & 32113.55 & 0.0020(3) & 0.0019(3) \\
	$12p_{1/2}$   & 0.016(1) & 32433.50 & 0.0012(2) & 0.0012(2) \\
	$(n>12)p_{1/2}$ tail   & & & 0.039(39) & 0.038(38) \\
	\hline
	Total $(n>8)p_{1/2}$ tail   & &  & 0.054(41) & 0.053(41) \\[0.3pc]
	%\hline
	$5p_{3/2}$    & 5.978(5) & 12816.55 & -83.824(126) & -84.937(129) \\
	$6p_{3/2}$    & 0.5230(11) & 23792.59 & 228.446(962) & 65.637(276) \\
	$7p_{3/2}$    & 0.202(4) & 27870.11 & 0.392(17) & 0.383(16) \\
	$8p_{3/2}$    & 0.111(3) & 29853.80 & 0.082(5) & 0.081(4) \\
	\hline
	$9p_{3/2}$    & 0.073(5) & 30970.19 & 0.0303(44) & 0.0299(43) \\
	$10p_{3/2}$   & 0.053(4) & 31661.16 & 0.0146(21) & 0.0144(21) \\
	$11p_{3/2}$   & 0.040(3) & 32118.52 & 0.0082(12) & 0.0081(12) \\
	$12p_{3/2}$   & 0.033(2) & 32437.04 & 0.0053(8) & 0.0052(8) \\
	$(n>12)p_{3/2}$ & & & 0.138(138) & 0.136(136) \\
	\hline
	Total $(n>8)p_{3/2}$ tail   & & & 0.196(146) & 0.194(144) \\[0.3pc]
	%\hline
	Core & & & 8.709(93) & 8.709(93) \\[0.3pc]
	Total $\alpha$ & & & 0.000 & 0.000 \\
    \end{tabular}
  \end{ruledtabular}
\end{table*}

The lowest order and all-order values of the ratio $R$ of the
$5s-6p_{3/2}$ and $5s-6p_{1/2}$
matrix elements are listed in Table~\ref{ratio}.  In the relativistic
all-order method, all
single-double (SD) or single-double and partial valence triple (SDpT)
excitations of the Dirac-Fock
(DF) wave function are included to all orders of perturbation
theory~\cite{relsd}. Additional
semi-empirical scaling of the all-order values was carried out for
both SD and SDpT values. The
scaled SD values (SD$_{\mathrm{sc}}$) is taken as final. The
uncertainty in the values of the ratio is
estimated  as the the maximum difference of the SD, SDpT,  and
SDpT$_{\mathrm{sc}}$ values from the
final result. The same procedure was used to determine theoretical
uncertainty in the values of
both matrix elements.

\begin{table*}
    \caption{The values of the $5s-6p_{j}$ reduced electric-dipole
	matrix elements in atomic
	units and their ratio $R$. The first-order (DHF) and all-order SD
	and SDpT values are listed; the
	label ``sc'' indicates the scaled values. \label{ratio}}
	\begin{ruledtabular}
      \begin{tabular} {lcccccr}
			  &  DHF& SD   &   SDpT          &  SD$_{\mathrm{sc}}$ &
			  SDpT$_{\mathrm{sc}}$ & Final\\
			    \hline
				$5s-6p_{1/2}$ &   0.3825&  0.3335  &0.3300 & 0.3248&
				0.3338&  0.325(9) \\
				$5s-6p_{3/2}$ &   0.6055 &0.5409 &0.5354 & 0.5276 &
				0.5401 & 0.528(13) \\
				$R$  &           1.5831& 1.6217& 1.6223&  1.6245&
				1.6179& 1.624(7)
      \end{tabular}
    \end{ruledtabular}
\end{table*}

\newpage
\end{widetext}

\end{document}